# Simulation of energy barrier distributions using real particle parameters and comparison with experimental obtained results

M Büttner[1], M Schiffler[2], P Weber[1] and P Seidel[1]

[1] Institut für Festkörperphysik, Friedrich-Schiller-Universität Jena, Helmholtzweg 5, 07743 Jena, Germany
[2] Institut für Geowissenschaften, Friedrich-Schiller-Universität Jena, Burgweg 11, 07749 Jena, Germany

Short Title: Simulation of energy barrier distributions

Corresponding author:

Markus Büttner, Friedrich-Schiller-Universität Jena,
Institut für Festkörperphysik, Helmholtzweg 5, 07743 Jena, Germany,
Email: Markus.Buettner@uni-jena.de
Tel.: ++49-3641-947387
Fax: ++49-3641-947412

## Abstract

In this work we compare previously measured energy barriers over the course of temperature with the results of simulations of the behaviour of the energy barriers. For the measurements the temperature dependent magnetorelaxation method (TMRX) was used. For the simulations of the energy barrier distribution we have used the real particles properties such as anisotropy and core size volume of the fractions of two magnetically fractionated ferrofluids. There is a good agreement between the simulated behaviour and the experimental obtained results. The influence of the particle volume concentration and agglomeration on the energy barrier distribution has been investigated. Finally the simulations confirm a previously published explanation for an experimentally obtained relaxation effect.

## 1. Introduction

The examination of the magnetic properties of many-particle systems of magnetic nanoparticles (MNP) is still a field of intensive research [1, 2]. One special method for the characterisation of MNP is based on the analysis of the temperature-dependent Néel-relaxation signal (TMRX). The theoretical mediations in this field of research are done by R.W. Chantrell et al. [3] and D. V. Berkov et al. [4-6]. Based on this theoretical work a system for the measurement of the temperature dependence of the Néel-relaxation signal were developed. In a first step measurements between room temperature and liquid nitrogen temperature were realised [7] and in a second step the temperature range were expanded down to liquid helium temperature [8]. With this measurement set-up a lot of interesting measurements were done [9, 10]. The missing connective link was the comparison between the results of a concrete simulation of the energy barriers using the real particle parameters and the experimental obtained results. In this work we compare previously published experimentally measured energy barrier distributions over the course of temperature [8, 10] with the calculated energy barrier distributions. For this comparison we use two varying ferrofluids that were magnetically fractionated to obtain smaller distribution of energy barriers and magnetically active core sizes, respectively.

## 2. Experimental details

A detailed description of the measurement system and the preparation procedure has been published elsewhere [8]. The first ferrofluid we use is a water-based ferrofluid (DDM 128N, Meito Sangyo, Japan) consisting of MNP with a core of iron oxide and a shell of carboxydextran. The second one also is a water-based ferrofluid (V190 [11]) consisting of MNP with a core of iron oxide but with a shell of carboxymethyldextran. The original ferrofluid with a broad particle size distribution was fractionated in an inhomogeneous magnetic field using an electromagnet with variable magnetic field strength (Bruker B-E 10v) and a magnetic separation column (MACS XS, Miltenyi Biotech) generating a strong field gradient. The particles are retained in the column in dependence on this field gradient as well as the particle magnetic moment. In what follows, the samples will be named with the used fractionation coil current. To obtain more reliable data we only use the fractionated samples but not the inititial solutions for our investigations. The magnetic data that is necessary for the simulation and the experimental obtained energy barrier distributions are taken from [10] for sample V190 und from [8] for sample DDM 128N.

## 3. Theoretical details

*3.1. Derivation of the important energy terms*

Our aim is to simulate the full shape of the energy barrier distribution found in TMRX measurements. The energy barrier distribution is the probability distribution to find an energy barrier to be overcome at one transition from an initial metastable state of magnetic nanoparticles to a final one. The energy barrier therefore is
$$\Delta E = E_{\max}(x_{opt}) - E_A, \qquad (1)$$
where $E_{\max}(x_{opt})$ is the maximum of the optimal path through the energy landscape which minimises the action as function of the magnetisation directions of the MNP and $E_A$ the initial energy of one metastable state.

According to [5, 12] we now have to find this optimal path via numerical minimization of the action. If the minimum condition $\nabla S \equiv 0$ is fulfilled, this optimal path should lead from the first metastable state over the saddle point to the second metastable state. There are two possible methods to find the minimum: The first one is the derivation the *N* Euler-Lagrange-equations from the action and solving the corresponding boundary value problem. For systems with e.g. *N*=128 MNP the inversion of the resulting equation system is time-consuming. For faster calculation the evaluation of the integral is better performed by numerical computation of the discretised variant of the action

$$S_{disc}(\Omega) = \Delta t \sum_{k=0}^{K-1} \sum_{i=1}^{N} \left[ \frac{\theta_{i,k+1} - \theta_{i,k}}{\Delta t} + \frac{1}{2}\left( \frac{\partial E\{\Omega_{k+1}\}}{\partial \theta_{i,k+1}} + \frac{\partial E\{\Omega_k\}}{\partial \theta_{i,k}} \right) \right]^2$$

$$+ \left[ \frac{\sin\theta_{i,k+1} + \sin\theta_{i,k}}{2} \frac{\phi_{i,k+1} - \phi_{i,k}}{\Delta t} + \frac{1}{2}\left( \frac{1}{\sin\theta_{i,k+1}} \frac{\partial E\{\Omega_{k+1}\}}{\partial \phi_{i,k+1}} + \frac{1}{\sin\theta_{i,k}} \frac{\partial E\{\Omega_k\}}{\partial \phi_{i,k}} \right) \right]^2 \quad (2)$$

for the set $\Omega$ of all magnetisation orientation angles $(\theta_{i,k}, \phi_{i,k})$. Via minimisation of this functional of all particle angles for the time slices $k = 1, \ldots, K-1$ with $k = 0$ and $k = K$ fixed representing the two metastable states we obtain the optimal trajectory. Thereby the system energy $E(\Omega)$ includes the anisotropy energy in reduced units

$$\varepsilon_{an} = -\frac{\beta}{2} \sum_i (\vec{m}_i \cdot \vec{n}_i)^2 \quad (3)$$

where ß is proportional to the anisotropy constant and $\vec{n}_i$ is the unit vector of the easy axis of each particle. The magnetisation unit vector $\vec{m}_i$ is achieved via dividing the magnetisation by its absolute value. Below the Curie temperature it can be identified with the saturation magnetisation of the system. The stray field energy is calculated with

$$\varepsilon_{dip} = -\frac{2\pi}{3} \sum_i \left( \sum_{r_{ij} < r_{rest}} \frac{3(\vec{e}_{ij} \cdot \vec{m}_i)(\vec{e}_{ij} \cdot \vec{m}_j) - \vec{m}_i \cdot \vec{m}_j}{r_{ij}^3} + \eta \vec{m}_i \langle m \rangle \right) \quad (4)$$

with the distance units rescaled by $R = r/a$ where $a$ denotes the average particle radius. To reduce calculation time significantly the Lorentz cavity method is used. The direct calculation of the stray field via the dipolar dependence acting on the *i*-th particle only is done inside a sphere around it with a restriction radius $r_{rest}$ twice the average interparticle distance. The remnant interactions are considered by the system magnetisation. The average system magnetisation is denoted by $\langle m \rangle$ in equation 4.

*3.2. Necessary simulation parameters*

Summarising the initialisation procedure for the simulation following parameters are needed for the run:

The reduced anisotropy constant $\beta = \frac{2K}{\mu_0 M_S^2}$ is required for calculation of equation 3 and can be obtained via magnetometric measurements. The anisotropy constant $K$ is the product of the coercivity and the saturation magnetisation: $K = H_C M_S$.

The particle concentration has to be calculated from the sample geometry and the individual average volume of the magnetic nanoparticles. If we denote the iron concentration with $n_{MNP}$, the molar mass $M_{MNP}$ and the density $\rho_{MNP}$ of the material the magnetic nanoparticles consist in and the sample volume with $V_{sample}$ we obtain:

$$\eta = \frac{n_{MNP} M_{MNP}}{\rho_{MNP} V_{sample}}. \quad (5)$$

The third important parameter is the number of runs of the minimisation procedure. Because the transition from one metastable state to another can be reverted we obtain two energy barriers in one run. With as many runs as possible we achieve statistical behavior of the system and therefore a more realistic energy barrier distribution. The latter expenses on calculation time: for getting 256 energy barriers we need one and a half weeks to perform full computation. Because of the complexity of the simulation program we are prevented to use powerful parallel computing techniques. Nevertheless multiprocessing can be done by running the program in several processes separately. By the way this workaround improves the stability of the whole simulation.

*3.3. Simulation procedure*

To find a certain energy barrier the simulation program has to do the following:
Firstly we need the initial configuration of the system of the MNP. It consists in the initial and final set of magnetisation directions describing a system of *N* single-domain particles. In our model we assume uniaxial single-particle anisotropy. To calculate the stray field between these particles we have to consider the spatial distribution of the MNP. The initialisation of the positions of these is done in an external script: The spatial matrix with 3*N* values representing *N* particles with three dimensions is obtained by using a random number generator and scaling by

$$2\left(\frac{N}{\eta}\right)^{1/3}, \tag{6}$$

where $\eta$ denotes the particle concentration. Factor 2 takes the scale of the spatial information to the average particle radius into account. The more particles are in the system is or the less the concentration is, so further the particles are divided. The spatial matrix also can be edited manually for achieving the possibility to let the simulation be influenced by the knowledge about the particles as much as possible. To obtain the initial configuration in magnetisation the search for the two metastable states has to be proceeded. There we have to find two energy minima via minimisation similar to the action minimiser. The main difference between the minimisation of the system consists in having only the first stage and calculating the values and gradients only of the energy. The break condition here is fulfilled, if the energy has undergone a target precision.

The action minimisation consists in three stages. In the first step several relaxation steps are made. This means the movement of the system coordinates from the initial configuration $\Omega^p$ into the direction of the local antigradient $\nabla S(\Omega^p)$ to the new position $\Omega^{p+1}$. According to

$$\Omega^{p+1} = \Omega^p - \alpha_{rel} \nabla S, \tag{7}$$

this shift is scaled by the step length $\alpha_{rel}$. This parameter is chosen arbitrarily, so that the action decreases. If after some step the action value increases we return to the previous configuration and let $\alpha_{rel} = \alpha_{rel}/2$. After ca. 10 relaxation steps the action should decrease after each step and therefore we can turn towards step two. In this stage the full function minimisation along the local antigradient is performed. With equation 7 we move the system coordinates with a step length of an optimised $\alpha_{opt}$ into the direction of the function minimum. In the third stage the gradient of the action has to be minimised. This step is to be used for determination of the relaxation step length $\alpha_{rel}$ at the new relaxation steps. Afterwards the program returns to step one and the relaxation steps are restarted. The break condition of the minimiser loop is achieved if for the components of the action gradient $\left|\partial S(\Omega^p)/\partial \theta_i\right| < \delta \forall i$ and $\left|\partial S(\Omega^p)/\partial \phi_i\right| < \delta \forall i$ is fulfilled. The target precision $\delta$ is an arbitrary constant, e.g. $\delta = 0.01$ leads to good results.

The described method for action minimisation seeks only for the shortest path over minimal energy values. Therefore it could happen that energy maxima are crossed by the "optimal" trajectory. To prevent this, a check procedure is enforced. So we move the system coordinates in the point with the maximum energy value into an arbitrary direction. The step length is chosen so that the new maximum gradient component at the new site is around ten times larger than the old one. The arbitrary moves of all other points are scaled with the average coordinate distance from one time step to another. With this configuration the minimisation procedure is resumed. If the second minimisation results in an energy value close to the first one we can surely assume that we have found an optimal trajectory leading over the saddle points of the energy landscape and the output is performed.

The result of the simulation program is a set of found energy barriers, the energy course in time, the corresponding action and the coordinates of the optimal trajectory. To obtain the energy barrier distribution we have to perform a simple last step manually. The distribution is nothing other than the histogram of the given energy barriers. To calculate the energy barrier distribution over temperature we have to rescale the system from the reduced values to the normal ones. According to

$$\varepsilon = \frac{E}{\mu_0 M_S^2 \overline{V}}, \tag{8}$$

there is to consider the scaling with the average particle volume $\overline{V}$. We can identify it with the maximum volume of the energy barrier density curves obtained from the TMRX measurements. The next step is to scale the information to one individual particle. Initially we observe the energy summarised for all N nanoparticles. Because the anisotropy energy determines the average height of the energy barrier this can be obtained by dividing the system energy by the expectation value of

$$\sum_i^n (\vec{m}_i \cdot \vec{n}_i)^2 = \sum_i^n \sin^2 \psi_i, \tag{9}$$

where $\psi_i$ denotes the angle between the easy axis and the magnetisation direction of each particle. The questioned value is nothing else than $N/2$. So the conversion factor of the energy between the reduced units and the measurement units is the described by

$$E = \frac{\mu_0 M_S^2 \overline{V}}{N/2} \varepsilon. \tag{10}$$

If we would like to get information about the temperature course in order to compare it with our TMRX measurements we have to divide the energy axis of the histogram by $20 k_B$ where the factor 20 is derived from the equation for the blocking volume [13] and $k_B$ is the Boltzmann constant.

## 4. Simulation results and discussion

The simulation program and the scaling described above were firstly run for two ferrofluids samples measured previously by TMRX. There are four interesting fractions of each sample prepared which are provided with information about the anisotropy constant obtained from magnetometric measurements. For the sample DDM 128N we obtain a particle volume concentration of $\eta = 0.000724$ using equation (6). Due to the mixture ratio between maghemite and magnetite in the ferrofluids system V190 this value is $\eta = 0.000814$. For each fraction 256 simulation runs are performed for finding a pair of energy barriers. The contribution of the anisotropy energy to the energy barrier is narrowly distributed around a center where the distribution of the concentration dependent stray field energy is broader. The simulations with a fit of the lognormal distribution to the histogram data are shown in figure 1.

Table 1 shows the simulation fit parameters for the investigated fractions of sample V190 and table 2 for the fractions of DDM 128N. The values for the anisotropy constant K, the volume of the particles $V_{max}$ and the experimental obtained temperature maximum of the energy barrier distribution $T_{maxExp}$ were taken from [8] for sample DDM 128N and from [10] for sample V190. The fit was done supposing a lognormal distribution of the particles. The factor ß in both tables denotes the calculated reduced anisotropy, $T_m$ the median and σ the standard deviation of the log-normal distribution. We firstly observe that the simulations reflect the lognormal distribution of the MNP found in real experiments. The differences between the peaks determined by simulation and fit on the one hand and the temperature detection deviation of the TMRX measurements on the other are caused by model-like character of the simulations and the error tolerance of the lognormal fit. To improve the amount of data two other simulations with the same systems of magnetic nanoparticles are performed. The first one should demonstrate the influence of the particle volume concentration on the shape of the energy barrier distribution (see fig. 2). This step is done for the 400 mA fraction of sample V190 and for the 0 A fraction of sample DDM 128N. It could be clearly shown that for very low concentrations $\eta = 0.0001$ the energy barrier closely is distributed around the expectation value of the anisotropy energy. Switching to higher concentrations ($\eta = 0.001$) for the sample DDM 128N we observe a broader distribution around $\varepsilon_{an}$. The broadening is less for the sample V190. A noticeable broadening for this sample starts at $\eta = 0.01$. For sample DDM 128N we observe a slightly shift of the maximum of the energy barrier distribution towards lower temperatures. For the sample V190 and $\eta = 0.01$ this is not applicable. For the highest simulated energy barrier distributions ($\eta = 0.05$) we observe in accordance to [4] a shift to lower temperatures and a broadening in the energy barrier distribution. Besides this, we also obtain energy barriers corresponding to higher temperatures. Due to the higher concentration the interparticle distance decreases. These increasing values of the stray field energy lead to higher energy values. After that a check with a system of partially agglomerated particles is done (see fig. 3). The agglomeration is done by compressing the spatial extension in all three dimensions for one fourth of the nanoparticles to the half of its value around the center of the system. That means, for a spatial part of $l^3 = \frac{1}{8}$ for the agglomerated particles their spatial information is scaled by $x \rightarrow \frac{1}{2}(1-l) + lx$. The investigations are performed for the same fractions described above. The results are shown in fig. 3. Therefore the energy barrier distribution broadens in comparison to the non-agglomerated investigations. For the sample V190 we only observe a slight shift to lower temperatures. For the fractions 250 mA and 0 mA of the sample DDM 128N the broadening is more significant. This could be lead back to the lower anisotropy constant. Figure 4 shows the comparison of the simulated energy barrier distribution over the course of temperature (bars) and the experimental obtained distribution (curves) for the fractions of sample V190 and DDM 128N. The experimental obtained energy barrier distributions are taken from [10] for sample V190 und from [8] for sample DDM 128N. As we see we are able to reconstruct the complete energy barrier distribution by only using average values of the (reduced) anisotropy constant and the average particle volume. The simulated energy barrier distributions over the course of temperature are in agreement with the measured ones. The simulations also show that the experimental obtained peak at 18 K for all fractions of the V190 sample is not a result of strongly interacting particles. Relaxation signals in an equal temperature range were also measured on magnetite nanoparticles produced by magnetotactic bacteria (so called magnetosomes) [9]. The origin of that signals was explained as memory effects [14]. This explanation can be confirmed by using the results of the simulations shown in figures 2 and 3.

**Conclusions**

The results of the simulations of the energy barrier distributions under use of the real particle parameters are in good agreement with the experimental obtained results. For very low particle concentrations the energy barrier closely is distributed around the expectation value of the anisotropy energy. In case of agglomeration the energy barrier distribution broadens in comparison to the non-agglomerated particle systems. The simulations reflect the lognormal distribution of the MNP found in real experiments and confirm a previously published explanation for an experimentally obtained relaxation effect which is not a result of the Néel relaxation process and therefore not an effect of the energy barrier.

**Acknowledgements**

The authors would like to thank D. V. Berkov for the fruitful discussions and the helpful suggestions.

**Table 1.** Parameters for the fractions of sample V190 used in simulation and for the fit.

| fraction [mA] | K [kJ/m³] | β | $V_{max}$ [$10^{-25}$m³] | $T_m$ [K] | σ | $T_{max\ Sim}$ [K] | $T_{max\ Exp}$ [K] | difference |
|---|---|---|---|---|---|---|---|---|
| 1000 | 13.7 | 0.18551 | 16.9 | 104.24 | 0.397 | 89.10 | 92 | -3.2% |
| 400 | 12 | 0.16303 | 28.2 | 167.95 | 0.437 | 138.72 | 150 | -7.5% |
| 200 | 13 | 0.17632 | 39.2 | 222.23 | 0.443 | 182.71 | 190 | -3.8% |
| 100 | 12.3 | 0.16713 | 46.0 | 272.28 | 0.410 | 230.06 | 225 | 2.2% |

**Table 2**. Parameters for the fractions of sample DDM 128N used in simulation and for the fit.

| fraction [mA] | K [kJ/m³] | β | $V_{max}$ [$10^{-25}$m³] | $T_m$ [K] | σ | $T_{max\ Sim}$ [K] | $T_{max\ Exp}$ [K] | difference |
|---|---|---|---|---|---|---|---|---|
| 6000 | 22.6 | 0.31120 | 0.61 | 7.58 | 0.242 | 7.15 | 5 | 43.0% |
| 1000 | 19.1 | 0.23101 | 1.99 | 16.83 | 0.287 | 15.51 | 13 | 19.3% |
| 250 | 11.4 | 0.15482 | 35.4 | 181.39 | 0.463 | 146.44 | 137 | 6.9% |
| 0 | 11.8 | 0.16225 | 56.4 | 322.71 | 0.414 | 271.93 | 210 | 29.5% |

**Figure captions**

**Figure 1.** Simulation with a fit of the lognormal distribution (curves) to the histogram data (bars) for the fractions of V190 (left) and DDM 128N (right).

**Figure 2.** Influence of the MNP concentration on the simulation result for the 400 mA fraction of sample V190 (left) and for the 0A fraction of sample DDM 128N (right). At low concentration there is only anisotropy energy.

**Figure 3.** Agglomeration of 25 percent of the particles for the fractions of sample V190 (left) and DDM 128N (right).

**Figure 4.** Comparison of the simulated energy barrier distribution over the course of temperature (bars) and the experimental obtained distribution (curves) for the fractions of sample V190 (left) and DDM 128N (right).

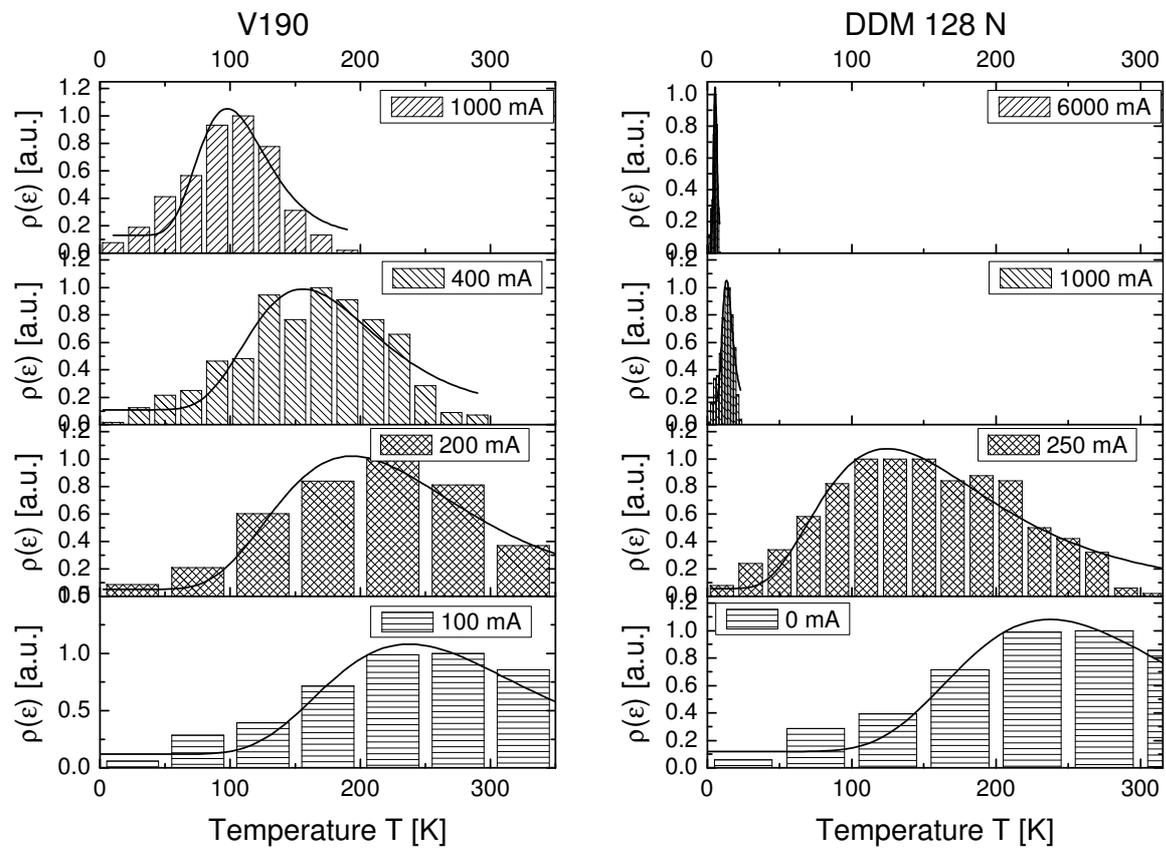

**Figure 1.** Simulation with a fit of the lognormal distribution (curves) to the histogram data (bars) for the fractions of V190 (left) and DDM 128N (right)

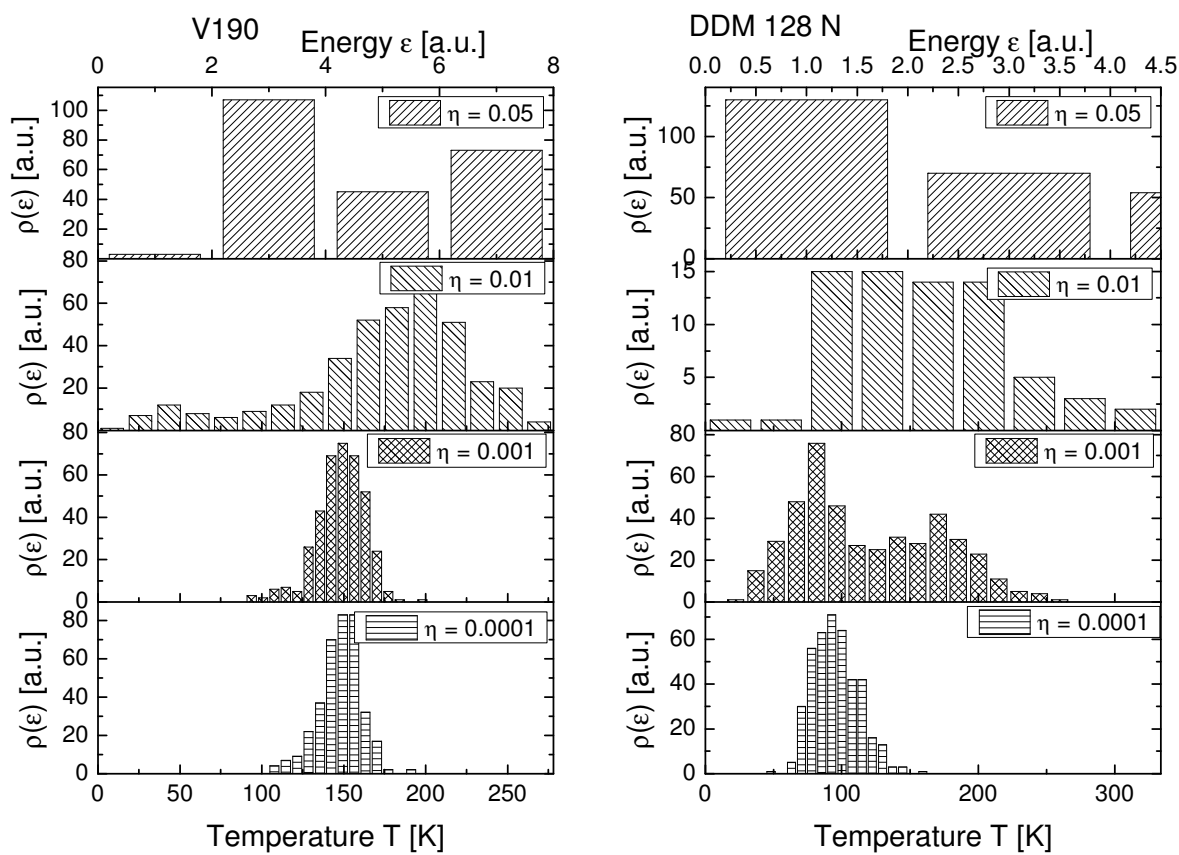

**Figure 2.** Influence of the MNP concentration on the simulation result for the 400 mA fraction of sample V190 (left) and for the 0A fraction of sample DDM 128N (right). At low concentration there is only anisotropy energy.

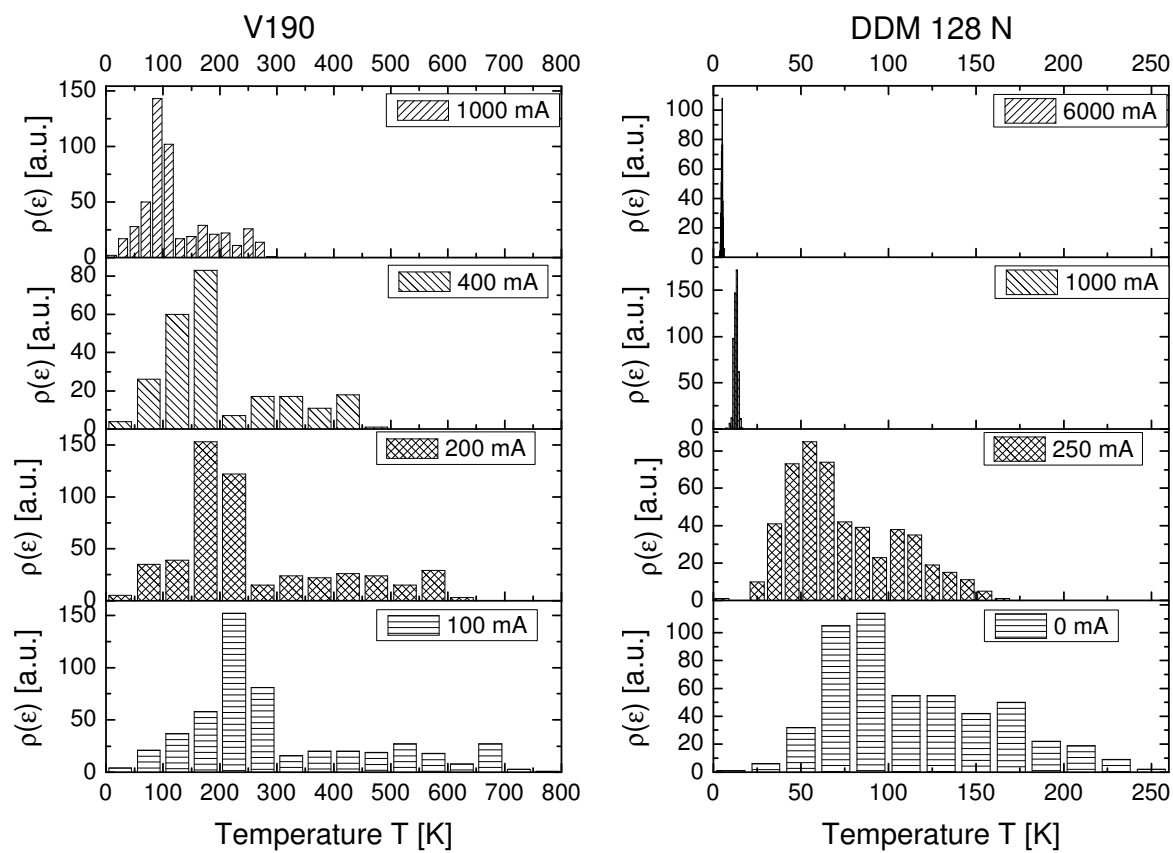

**Figure 3.** Agglomeration of 25 percent of the particles for the fractions of sample V190 (left) and DDM 128N (right).

**Figure 4.** Comparison of the simulated energy barrier distribution over the course of temperature (bars) and the experimental obtained distribution (curves) for the fractions of sample V190 (left) and DDM 128N (right).

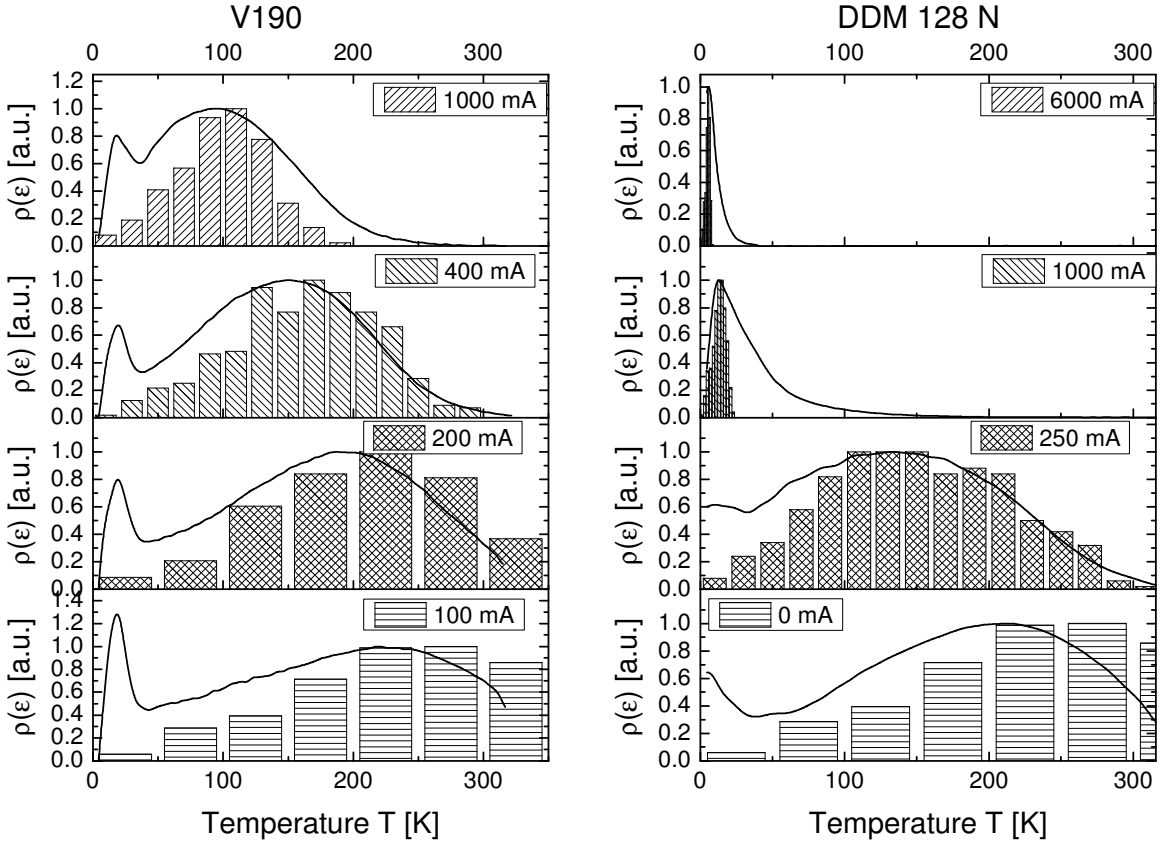